\documentclass[a4paper,twoside,english,british]{article}
\usepackage[T1]{fontenc}
\usepackage[latin9]{inputenc}
\usepackage{graphicx}

\makeatletter

\pdfpageheight\paperheight
\pdfpagewidth\paperwidth

\makeatother

\usepackage{babel}
\begin{document}

\title{\noindent {\Large Why is GDP growth linear?}}

\author{\noindent Jörg D. Becker \\
{\small e-mail: joerg.becker@icas-evolutions.de}}
\maketitle
\begin{abstract}
\textbf{In many European countries the growth of the }\textbf{\emph{real}}\textbf{
GDP }\textbf{\emph{per capita}}\textbf{ has been linear since 1950.
An explanation for this linearity is still missing. We propose that
in artificial intelligence we may find models for a linear growth
of performance. We also discuss possible consequences of the fact
that in systems with linear growth the percentage growth goes to zero.}
\end{abstract}
In Babylonia the \emph{real} GDP \emph{per capita} has shown a linear
growth from 1950 up to now (2015), as in practically all European
countries (up to fluctuations) {[}1,2,3,4,5{]}. In the USA growth
has indeed been exponential, but this is due to the exponential growth
of the population: More people eat more and work more. There, the
\emph{real per capita} GDP has grown linearly, too. Of course, linear
growth may break down due to a particular incidence, as it happened
to Russia in 1989.

\begin{figure}[bh]
\noindent \begin{centering}
\includegraphics[scale=0.5]{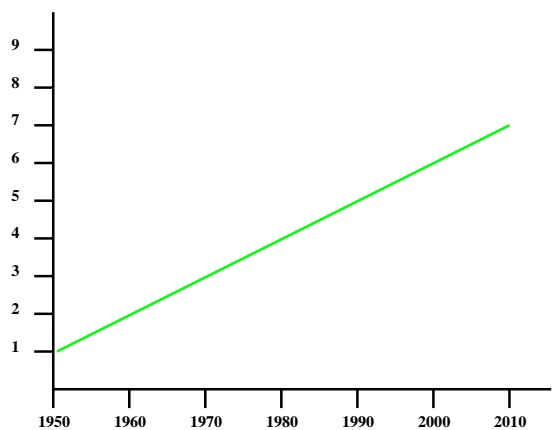}
\par\end{centering}

\caption{Growth of \emph{real} GDP \emph{per capita} in Babylonia (in multiples
of the GDP in 1950)}
\end{figure}

How could such a linear growth process start? And what is the mechanism
behind it?

For a physicist, among all possible types of lines - exponential,
logarithmic, logistic, wavelike, chaotic, stochastic - a straight
line is indeed something very special and thus calls for an explanation.

So far I don't know any answer to this question. Some economists say
that it is exponential but the second term is so small that we don't
see it yet; others say that this linearity is purely accidental; still
others say that linear growth cannot occur, so we need not talk about
it.

Sure, there are systems with an exponential growth: compound interest,
bacteria in a Petri dish, Moore's Law; others start with an exponential
growth and then run into saturation (logistic growth): the number
of Gothic cathedrals in Europe, the number of burned witches in Scotland,
and the $CO_{2}$ concentration in the atmosphere {[}6,7,8,11{]}.
These two types are related to mathematical models: The equation $dB/dt=\alpha\cdot B$
leads to exponential growth, and $dB/dt=\alpha\cdot B-\beta\cdot B^{2}$
leads to logistic growth (with $\alpha$ and $\beta$ system specific
parameters).

For a linear growth we may take the bathtub as an example, with a
constant flow of water, which could be modeled by the equation $dB/dt=\alpha$
. However, I find it hard to believe that economy could be seen as
a system just nourished by a constant flow of something.

Something more sophisticated may be needed for an explanation. An
example is the use of thermodynamic models in economics, but as far
as I know also this approach cannot reproduce a linear growth of the
GDP.

However, in artificial intelligence one could possibly meet systems
with such a behaviour, like in some Pseudo-Monte-Carlo systems. Here,
in a complex learning system, at a random spot some change is applied,
and if the performance of the system is increased the change will
be kept, otherwise it will be reversed; and this proceedure will be
repeated at some other spot etc. In such a system the performance
may grow linearly.

This could serve as a possible model for economics: For instance,
at some spot an engineer modifies a diesel engine, and if this leads
to a better performance the modification will be kept; or if a company
brings to market a visual telephone, and this turns out to be a flop,
then the company will stop producing these devices. (Later on the
idea of a visual telephone may successfully be implemented in a different
way.) Actually, if progress in economy is based on many single logistic
processes taking place in time, the sum of these may have a linear
line as an evelope; but this is purely speculative.

In artificial intelligence there are other processes which are linear
in time. In ``sane'' artificial evolution the pragmatic information
grows linearly in time (up to fluctuations), but in this case the
information is the logarithm of the performance which itself is exponential.
Now we could turn the argument around and say that there is some economic
quantitiy X which grows exponentially, and the GDP is just its logarithm.
However, this would somehow contradict the fact that the pragmatic
information contained in a quantity of money is proportional to its
logarithm {[}11{]}. 

Let us now consider more closely some systems with logistic growth
{[}6,7,8{]}. The construction of Gothic cathedrals is an example.
In those times the population in the cities was growing very fast,
so there was a need for larger churches, and we may call this a technical
reason for growth. On the other hand, Gothic cathedrals were in fashion,
which can be seen as a ``cultural program'', a term which (in this
context) has been coined by Cesare Marchetti.

In the case of burning witches we can hardly speak of a technical
process, because at that time the technology of stakes was well developed;
so this seems to have been a purely cultural program. (See also {[}9{]}.)

It is often difficult to distinguish between technical and cultural
processes. Consider for example Moore's Law which says that the number
of transistors on a chip doubles every 18 months. Thus the relative
growth rate is a constant leading to an exponential growth. Is this
a technical or a cultural process? At first we may say that there
are technical reasons for this growth behaviour, like the development
of technology or the money available for R\&D; but we may also suspect
that it is a cultural program or even a cultural pressure such that
the protagonists - physicists, engineers, managers - act in a way
to fulfill the program.

In the case of GDP growth we are faced with the same question: Is
there a technical reason for linear growth, or is this a cultural
process such that the protagonists - engineers, managers, politicians,
jurists - act in a way to fulfill the program of linear growth? In
the latter case we may say that somewhere there is some kind of information
in the system which is preserved and determines the process.

Another amazing fact is that economic growth has been remarkably stable
against the political direction of the government, against the social
policy, against several tax reforms, against the distribution of wealth,
against external disturbances (the influence of the two oil crises,
e.g., was very small indeed), with the gobalisation, with innovations
like the internet, etc.

An unavoidable consequence of linear growth is that the relative growth
rate itself decreases. Clearly if $B$ grows linearly, $B=\alpha$$\cdot t$
, the growth rate $dB/dt=\alpha$ is a constant, the \emph{relative}
growth rate $r=1/B\cdot dB/dt=1/(\alpha\cdot t)\cdot\alpha$ goes
to zero like $1/t$ , and the same of course applies to the percentage
growth rate which is just $g=100\cdot r$. So, for instance, globalization
did not lead to a faster growth of economy; but we cannot exclude
that it was a measure to keep the slope $\alpha$ of growth constant.

Thus, whereas in the decade from 1950 to 1960 the GDP increased by
100 \%, the percentage growth went down to an increase of 17 \% between
2000 and 2010, corresponding to an average annual percentage growth
rate of 1.7 \%; now, in 2015, the percentage growth rate is reported
to be around 1.3 \% (with fluctuations, of course).

In the seventies people thought that a stable economy without growth
was not possible. Now we have an experimental proof that a (nearly)
steady-state economy is indeed possible (unless you claim that it
is not stable).

Is there a connection between the growth rate and the level of interest?
And is the presently low interest rate necessarily smaller than the
percentage growth rate? We may assume that interest can only be generated
by a surplus in the production. Then it cannot be larger than the
growth rate of economy. Actually we see that the interest produced
by a life insurance can no longer reach values of 4\% or larger; and
one of the problems connected to the Euro Stability Pact may be due
to the assumption that the percentage growth rate of the debts may
achieve up to $g=3\%$ (a number connected to the Holy Trinity), whereas
now the percentage growth rate of the GDP has become much smaller.
So we might suspect that the decline of the relative growth rate like
$g\sim1/t$ was not taken into account when the Euro Stability Pact
was closed. We also have to keep this in mind when talking about the
crisis connected to Greece, and the interest rates they have to pay
for their debts.

Finally we may ask why investigations remain low in spite of the fact
that loans are cheap. Could it be that also industry does not believe
in a larger growth rate in demand? And how could we possibly start
a new cycle of growth (linear or other)? Or is this unnessecary?

Remark: Some if these questions have already been discussed in references
{[}10,11{]}.

\subsubsection*{Acknowledgment:}

It is a pleasure to thank Dr. Kay Bourcarde for his hint to ref. {[}3{]}.

\subsubsection*{References:}

{[}1{]} Horst Afheld (1997), \foreignlanguage{english}{Wohlstand für
Niemand? Die Marktwirtschaft entläßt ihre Kinder. Rohwolt Tb. ISBN:
978-3499604720 }

{[}2{]} Kernaussage des Instituts für Wachstumsstudien (Edition 2013).
Dr. Kay Bourcarde, Institut für Wachstumsstudien, Postfach 11 12 31,
D-35357 Gießen 

{[}3{]} Sören Wibe, Ola Carlén, Is Post-War Economic Growth Exponential?
Australian Economic Review, Vol. 39, No. 2, pp. 147-156, June 2006

{[}4{]} Carlén Ola, Wibe Sören (2012), A Note on Post-War Economic
Growth. The Empirical Economics Letters, 11 (2)

{[}5{]} http://antipropaganda.eu/USSRGrowth.html

\selectlanguage{english}%
{[}6{]} Cesare Marchetti (1996), Looking Forward \textendash{} Looking
Backward. A Very Simple Mathematical Model for Very Complex Social
Systems. To be found in:\\
http://cesaremarchetti.org/publist.php

{[}7{]} Cesare Marchetti, A Personal Memoir: From Terawatts to Witches.
My Life with Logistics at IIASA. Technonlogical Forecasting and Social
Change 37, 409-414 (1990)

{[}8{]} John Casti, Reality Rules, The Fundamentals; in particular:
p. 131. Wiley Interscience (1990) ISBN: 978-0471570219

{[}9{]} Christopher Fry, The Lady's Not for Burning. Dramatist's Play
Service (January 1998). ISBN: 0-8222-1431-8

\selectlanguage{british}%
{[}10{]} J.D.Becker (Hrsg), Wirtschaftsphysik (2 Bände). Proceedings
eines Workshops, München, September 2008. Universität der Bundeswehr
München 2008 

{[}11{]} Jörg D. Becker (2014), Information Theory - A Tool for Thinking.
Shaker Verlag. ISBN: 978-3-8440-2719-8 
\end{document}